\documentclass[12pt,a4paper]{article}
\usepackage{mathrsfs}
\usepackage{epsfig}
\begin{document}
\title{Single production of the top partners \\ at high energy
colliders}
\author{Chong-Xing Yue, Hui-Di Yang, Wei Ma\\
{\small Department of Physics, Liaoning  Normal University, Dalian
116029, P. R. China}
\thanks{E-mail:cxyue@lnnu.edu.cn}}
\date{\today}

\maketitle
\begin{abstract}

The left-right twin $Higgs$ ($LRTH$) model is a concrete realization
of the twin $Higgs$ mechanism, which predicts the existence of the
top partner $T$. In this paper, we consider production of $T$
associated with the top quark $t$ at the high energy linear
$e^{+}e^{-}$ collider ($ILC$) and the $LHC$ experiments, and its
single production in future linac-ring type $ep$ collider
experiment. To compare our results with those of the littlest
$Higgs$ model with $T$-parity, we also estimate production of the
$T$-even top partner $T_{+}$ via the corresponding processes in
these high energy collider experiments. A simply phenomenological
analysis is also given.

\vspace{1cm}

\end{abstract}
\newpage
\noindent{\bf 1. Introduction}

Although the standard model ($SM$) of elementary particle physics
provides a very successful description of existing experiments at
the highest energies currently accessible at colliders, there is the
fine tuning problem to be solved in the $SM$. To solve the fine
tuning problem, many new physics models beyond the $SM$ have been
proposed. However, a so-called " little hierarchy problem " [1]
arises in these models, once constraints from precision measurements
are imposed. So far, there are some successful models solving the
little hierarchy problem, for example the $MSSM$ with $R$-parity
[2], the little $Higgs$ models [3], the universal extra dimensional
model with $KK$-parity [4], or the twin $Higgs$ mechanism [5].

It is well known that the top loop in the $SM$ is the largest
contribution to the $Higgs$ mass quadratic divergence. Thus, for the
new physics models to solve the fine tuning problem, there must be
some particles constrained by symmetry, which cancel this loop. All
of the models mentioned above contain a heavy particle which shares
the gauge quantum numbers of the top quark, generally called " top
partner "[6]. This kind of new particles can lead to a relatively
generic class of collider signals from their production and
subsequent decay, which have been extensively studied in the
literature [7, 8, 9].

Recently, the twin $Higgs$ mechanism has been proposed as a solution
to the little hierarchy problem [5]. The twin $Higgs$ mechanism
proceeds in two main steps: i) the $SM$ $Higgs$ emerges as a
pseudo-$Goldstone$ boson from a spontaneously broken global
symmetry; ii) an additional discrete symmetry is imposed, which can
make that the leading quadratically divergent terms cancel each
other and do not contribute the $Higgs$ mass. The twin $Higgs$
mechanism can be implemented in left-right models with the
additional discrete symmetry being identified with left-right
symmetry [10, 11]. The left-right twin $Higgs$ ($LRTH$) model
contains the $U(4)_{1} \times U(4)_{2}$ global symmetry as well as
the gauged symmetry $SU(2)_{L} \times SU(2)_{R} \times U(1)_{B-L}$.
After $Higgs$ obtain vacuum expectation values, the global symmetry
$U(4)_{1} \times U(4)_{2}$ breaks down to $U(3)_{1} \times
U(3)_{2}$, and the gauge group $SU(2)_{R} \times U(1)_{B-L}$ breaks
down to the $SM$ $U(1)_{Y}$. The leading quadratically divergent
contributions of the $SM$ gauge bosons to the $Higgs$ mass are
canceled by the loop involving the new gauge bosons, while those for
the top quark can be canceled by the contributions from a heavy top
quark called top partner. Thus, the $LRTH$ model predicts the
existence of the new particles, such as heavy gauge bosons, heavy
scalars, and the top partner, which can generate rich phenomenology
at present and in future collider experiments [11, 12, 13, 14].

The single and pair production of the top partners  predicted by the
$LRTH$ model at the $LHC$ are studied in Ref. [11]. As we know, so
far, in the context of the $LRTH$ model, production of the top
partner associated with the top quark $t$ has not been considered at
the high energy linear $e^{+}e^{-}$ collider ($ILC$), the linac-ring
type $ep$ collider ($THERA$), and the $LHC$, which is the main aim
of this paper. There are several motivations to perform this study.
First, so far, most of works about the top partner focus on
phenomenology analysis at the $LHC$. Studies about production of the
top partner at $ILC$ and $THERA$ are very few. Second, the top
partner predicted by different new physics models might generate
similar signatures at the $LHC$. It is very difficult to
differentiate each other. Third, as long as the top partner is not
too heavy, it can be singly produced in future $ILC$ and $THERA$
experiments. These high energy collider experiments with more clear
environment could help us to distinguish different new physics
models. Thus, in this paper, we will completely consider single
production of the top partner in these three kinds of the high
energy collider experiments and compare the numerical results with
each other.

The layout of the present paper is as follows: In section 2, we
briefly review the essential features of the $LRTH$ model. The
relevant couplings of the top partner $T$ to other particles are
also given in this section. Single production of $T$ in future $ILC$
, $THERA$, and $LHC$  experiments are calculated in sections 3, 4,
and 5, respectively. In these sections, we discuss the possible
signals generated by the processes $e^{+}e^{-} \rightarrow
t\bar{T}+\bar{t}T$, $ep \rightarrow \nu_{e}\bar{T }+ X$, and $pp
\rightarrow t\bar{T}+\bar{t}T + X$. To compare our results obtained
in the context of the $LRTH$ model with those of the littlest
$Higgs$ model with $T$-parity, called the $LHT$ model [15], we
further consider production of the $T$-even top partner $T_{+}$ via
the corresponding processes in these three sections. Finally, our
conclusions are given in section 6.

\noindent{\bf 2. The $LRTH$ model}

The $LRTH$ model was first proposed in Ref. [10] and the details of
the model as well as the particle spectrum, $Feynman$ rules, and
some phenomenology analysis have been studied in Ref. [11]. Here we
will briefly review the essential features of the model and focus
our attention on the heavy gauge bosons and heavy top partner.

The $LRTH$ model is based on the global $U(4)_{1} \times U(4)_{2}$
symmetry with a locally gauged subgroup $SU(2)_{L} \times SU(2)_{R}
\times U(1)_{B-L}$. Two $Higgs$ fields, $H = (H_{L}, H_{R})$ and
$\hat{H} = (\hat{H}_{L}, \hat{H}_{R})$, are introduced and each
transforms as $(4, 1)$ and $(1, 4)$ respectively under the global
symmetry. $H_{L, R}$ ($\hat{H}_{L, R}$) are two component objects
which are charged under $SU(2)_{L}$ and $SU(2)_{R}$, respectively.
For the gauge couplings $g_{2L}$ and $g_{2R}$ of $SU(2)_{L}$ and
$SU(2)_{R}$, the left-right symmetry implies that $g_{2L}$ =
$g_{2R}$ = $g_{2}$.

The $U(4)_{1}$ ($U(4)_{2}$) group is spontaneously broken down to
its subgroup $U(3)_{1}$ ($U(3)_{2}$) with non-zero vacuum
expectation value ($VEV$) $< H >$ = ($0$, $0$, $0$, $f$) ($< \hat{H}
>$ = ($0$, $0$, $0$, $\hat{f}$)). The $Higgs$ $VEVs$ also break
$SU(2)_{R} \times U(1)_{B-L}$ down to the $SM$ $U(1)_{Y}$. After
spontaneous global symmetry breaking by $f$ and $\hat{f}$, three
$Goldstone$ bosons are eaten by the new gauge bosons $W_{H}^{\pm}$
and $Z_{H}$. After the electroweak symmetry breaking, the three
additional $Goldstone$ bosons are eaten by the $SM$ gauge bosons
$W^{\pm}$ and $Z$. In the $LRTH$ model, the masses of the heavy
gauge bosons can be written as:
\begin{equation}
M_{W_{H}}^{2} = \frac{1}{2}g_{2}^{2}(\hat{f}^{2}+f^{2}\cos^{2}x),
\end{equation}
\begin{eqnarray}
M_{Z_{H}}^{2} =
\frac{g_{1}^{2}+g_{2}^{2}}{g_{2}^{2}}(M_{W_{H}}^{2}+M_{W}^{2})-M_{Z}^{2},
\end{eqnarray}
where $x=v/\sqrt{2}f$. $g_{1}$ and $g_{2}$ (= $g_{2L}$ = $g_{2R}$)
are the gauge coupling constants of the $U(1)_{B-L}$ and $SU(2)_{L,
R}$, respectively, which can be written as:
\begin{eqnarray}
g_{1}=\frac{e}{\sqrt{\cos2\theta_{W}}},
\hspace{0.5cm}g_{2}=\frac{e}{S_{W}}.
\end{eqnarray}
Where $S_{W}= \sin\theta_{W}$ and $\theta_{W}$ is the $Weinberg$
angle.

The fermion sector of the $LRTH$ model is similar to that of the
$SM$, with the right handed quarks ($u_{R}$, $d_{R}$) and leptons
($l_{R}$, $v_{R}$) form fundamental representations of $SU(2)_{R}$.
In order to give the top quark mass of the order of the electroweak
scale, a pair of vector-like quarks $Q_{L}$ and $Q_{R}$ are
introduced. The mass eigenstates, which contain one the $SM$ top
quark $t$ and a heavy top partner $T$, are mixtures of the gauge
eigenstates. Their masses are given by
\begin{eqnarray}
m_{t}^{2} = \frac{1}{2}(M^{2}+y^{2}f^{2}-N_{t}),\hspace{0.5cm}
M_{T}^{2} = \frac{1}{2}(M^{2}+y^{2}f^{2}+N_{t}),
\end{eqnarray}
where $N_{t} = \sqrt{(y^{2}f^{2}+M^{2})^{2}-y^{4}f^{4}\sin^{2}2x}$.
Provided $M_{T} \leq f$ and that the parameter $y$ is of order one,
the top $Yukawa$ coupling will also be of order one. The parameter
$M$ is essential to the mixing between the $SM$ top quark and its
partner. At the leading order of $1/f$, the mixing angles can be
written as:
\begin{eqnarray}
S_{L}=\sin\alpha_{L}\simeq\frac{M}{M_{T}}\sin x,\hspace{0.5cm}
S_{R}=\sin\alpha_{R}\simeq\frac{M}{M_{T}}(1+\sin^{2}x).
\end{eqnarray}

The left(right) coupling constants of the gauge bosons and the top
partner $T$ to other particles, which are related our calculation,
can be written as:
\begin{eqnarray}
g_{L}^{Zt\bar{T}} = \frac{eC_{L}S_{L}}{2C_{W}S_{W}},\hspace{0.5cm}
g_{R}^{Zt\bar{T}} =
\frac{ef^{2}x^{2}S_{W}C_{R}S_{R}}{2\hat{f}^{2}C_{W}^{3}};
\end{eqnarray}
\begin{eqnarray}
g_{L}^{Ze^{+}e^{-}} =
\frac{e(-\frac{1}{2}+S_{W}^{2})}{S_{W}C_{W}},\hspace{0.5cm}
g_{R}^{Ze^{+}e^{-}} = \frac{eS_{W}}{C_{W}};
\end{eqnarray}
\begin{eqnarray}
g_{L}^{Z_{H}t\bar{T}} = \frac{eC_{L}S_{L}S_{W}}{2C_{W}\sqrt{\cos
2\theta_{W}}},\hspace{0.5cm} g_{R}^{Z_{H}t\bar{T}} =
-\frac{eC_{R}S_{R}C_{W}}{2S_{W}\sqrt{\cos 2\theta_{W}}};
\end{eqnarray}
\begin{eqnarray}
g_{L}^{Z_{H}e^{+}e^{-}} = \frac{2eS_{W}}{4C_{W}\sqrt{\cos
2\theta_{W}}},\hspace{0.5cm} g_{R}^{Z_{H}e^{+}e^{-}} =
\frac{e(1-3\cos 2\theta_{W})}{4S_{W}C_{W}\sqrt{\cos 2\theta_{W}}};
\end{eqnarray}
\begin{eqnarray}
g_{L}^{W\bar{T}b} =
\frac{eS_{L}}{\sqrt{2}S_{W}},\hspace{0.5cm}g_{L}^{W\nu_{e}e} =
\frac{e}{\sqrt{2}S_{W}};
\end{eqnarray}
\begin{eqnarray}
g_{R}^{W_{H}\bar{T}b} =
\frac{eC_{R}}{\sqrt{2}S_{W}},\hspace{0.5cm}g_{R}^{W_{H}\nu_{e}e} =
\frac{e}{\sqrt{2}S_{W}}.
\end{eqnarray}
Where $C_{L}^2=1-S_{L}^2$, $C_{R}^2=1-S_{R}^2$.

According the symmetry breaking pattern discussed above, with
certain reparametrization of the fields, there are left with four
scalars in the $LRTH$ spectrum that couple to both the fermion
sector and the gauge boson sector. They are one neutral pseudo
scalar $\phi^{0}$, a pair of charged scalars $\phi^{\pm}$, and the
$SM$ physical $Higgs$ $h$. In addition, there is an $SU(2)_{L}$
doublet $\hat{h} = (\hat{h}_{1}^{+}, h_{2}^{0})$ that couples to the
gauge boson sector only. It has been shown that the lightest
particle in $\hat{h}$, typically one of the neutral components, is
stable, and therefore constitutes a good dark matter candidate [14].
Thus, the top quark $t$ and its partner $T$ can couple to some of
these scalars:
\begin{eqnarray}
ht \bar{t} : - \frac{e}{2S_{W}}\frac{m_{t}C_{L}C_{R}}{m_{W}},
\hspace{0.5cm} hT \bar{T} : -
\frac{y}{\sqrt{2}}(S_{R}S_{L}-C_{L}C_{R}x);
\end{eqnarray}
\begin{eqnarray}
\phi^{0}t\bar{t} : -\frac{iy}{\sqrt{2}}S_{R}S_{L},
\hspace{0.5cm}\phi^{0}T\bar{T} : -\frac{iy}{\sqrt{2}}C_{L}C_{R};
\end{eqnarray}
\begin{eqnarray}
h\bar{T}t :
-\frac{y}{\sqrt{2}}[(C_{L}S_{R}+S_{L}C_{R}x)P_{L}+(C_{L}S_{R}x+S_{L}C_{R})P_{R}];
\end{eqnarray}
\begin{eqnarray}
\phi^{0}\bar{T}t :
-\frac{iy}{\sqrt{2}}[S_{L}C_{R}P_{L}-C_{L}S_{R}P_{R}],\hspace{0.5cm}
\phi^{+}\bar{T}b : \frac{i}{f}[C_{R}m_{b}P_{L} - yC_{L}fP_{R}].
\end{eqnarray}
Thus, the possible decay modes of the top partner $T$ are
$\phi^{+}b$, $W^{+}b$, $th$, $tZ$, and $t\phi^{0}$, which are
extensively studied in Ref. [11].

In the following sections, we will use the above $Feynman$ rules to
calculate the single production cross sections of the top partner
$T$ and discuss its possible signals at the $ILC$, $THERA$, and
$LHC$ experiments.

\noindent{\bf 3. Production of the top partner $T$ associated with
the top quark $t$ at the $ILC$}

From the above discussions, we can see that production of the top
partner $T$ associated with the top quark $t$ via $e^{+}e^{-}$
collision proceeds through the $S$-channel $Z$ exchange and $Z_{H}$
exchange as shown in Fig.1. Using Eqs.(6)-(9), the production cross
section $\sigma_{1}$ can be written as:
\begin{eqnarray}
\sigma_{1}(s)\hspace{0.5cm}=&&\frac{3
\sqrt{(s+M_{T}^{2}-m_{t}^{2})^{2}-4sM_{T}^{2}}}{8\pi s^{2}
}\{[\frac{((g_{L}^{Ze^{+}e^{-}})^{2}+(g_{R}^{Ze^{+}e^{-}})^{2})
((g_{L}^{Zt\bar{T}})^{2}+(g_{R}^{Zt\bar{T}})^{2})}{2(s-M_{Z}^{2})^{2}}
\nonumber\\&&
+\frac{((g_{L}^{Z_{H}e^{+}e^{-}})^{2}+(g_{R}^{Z_{H}e^{+}e^{-}})^{2})(
(g_{L}^{Z_{H}t\bar{T}})^{2}+(g_{R}^{Z_{H}t\bar{T}})^{2})}{2(s-
M_{Z_{H}}^{2})^{2}}\nonumber\\
&&+\frac{(g_{L}^{Ze^{+}e^{-}}g_{L}^{Z_{H}e^{+}e^{-}}+g_{R}^{Ze^{+}e^{-}}
g_{R}^{Z_{H}e^{+}e^{-}})
(g_{L}^{Zt\bar{T}}g_{L}^{Z_{H}t\bar{T}}+g_{R}^{Zt\bar{T}}g_{R}^{Z_{H}
t\bar{T}})}
{(s-M_{Z}^{2})(s-M_{Z_{H}}^{2})}]\nonumber\\&&(\frac{s^{2}-(M_{T}^{2}
-m_{t}^{2})^{2}}{4}+
\frac{(s+M_{T}^{2}-m_{t}^{2})^{2}-4sM_{T}^{2}}{12})\nonumber\\&&+
[\frac{g_{R}^{Zt\bar{T}}
g_{L}^{Zt\bar{T}}((g_{L}^{Ze^{+}e^{-}})^{2}+(g_{R}^{Ze^{+}e^{-}})^{2})
}{(s-M_{ Z}^{2})^{2}}+\frac{g_{R}^{Z_{H}t\bar{T}}
g_{L}^{Z_{H}t\bar{T}}((g_{L}^{Z_{H}e^{+}e^{-}})^{2}+(g_{R}^{Z_{H}
e^{+}e^{-}})^{2})}{(s-M_{Z_{H}}^{2})^{2}}\nonumber\\&&+\frac{(g_{L}
^{Ze^{+}e^{-}}g_{L}^{Z_{H}e^{+}e^{-}}+g_{R}^{Ze^{+}e^{-}}g_{R}^{Z_{H}e^{+}e^{-}})
(g_{R}^{Zt\bar{T}}g_{L}^{Z_{H}t\bar{T}}+g_{L}^{Zt\bar{T}}g_{R}^{Z_{H}t\bar{T}})}
{(s-M_{Z}^{2})(s-M_{Z_{H}}^{2})} ]m_{t}M_{T}s\}.
\end{eqnarray}
Where $\sqrt{s}$ is the center-of-mass ($c. m.$) energy of the
$ILC$.

\begin{figure}[htb]
\begin{center}
\epsfig{file=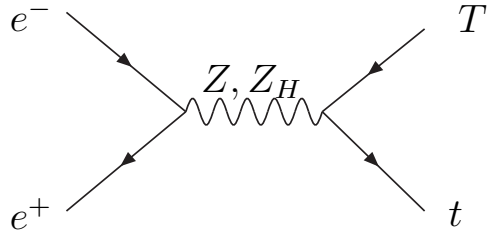,width=180pt,height=90pt}
 \caption{$Feynman$ diagrams for the process $e^{+}e^{-}\rightarrow\ t\bar{T}$ in the $LRTH$ model.}

\end{center}
\end{figure}

From the above expression, we can see that, except for the $SM$
input parameters $\alpha_{e}=1/128.8$, $S_{W}^{2}=0.2315$,
$m_{t}=172.5$ $GeV$, and $M_{Z}=91.187$ $GeV$ [16], the production
cross section $\sigma_{1}$ for the top partner $T$ at the $ILC$ is
dependent on the model dependent parameters $f$, $\hat{f}$, $M$, and
$M_{T}$ (or $y$). Once $f$ is fixed, the scalar $VEV$ $\hat{f}$ can
be determined by minimizing the $Coleman-Weinberg$ ($CW$) potential
for the $SM$ $Higgs$ boson and requiring that the $SM$ $Higgs$ boson
obtains an electroweak symmetry breaking $VEV$ of 246 $GeV$ [11].
The top $Yukawa$ coupling constant $y$ can be determined by fitting
the experimental value of the top mass $m_{t}$. The free parameters
$f$ and $M$ are constrained by the precision measurements. In our
numerical estimation below, we will assume $M\leq300$ $GeV$ and
$f\leq1500$ $GeV$.

Our numerical results are summarized in Fig.2, in which we plot
$\sigma_{1}^{LR}$ as a function of the $VEV$ value $f$ for the $c.
m.$ energy $\sqrt{s} = 2$ $TeV$ and three values of the mixing
parameter $M$. One can see from Fig.2 that the cross section
$\sigma_{1}^{LR}$ is very sensitive to the parameters $f$ and $M$.
When $M$ inclines to zero, its value goes to zero. This is because
$M=0$ leads no-mixing between the $SM$ top and its partner, the
$Zt\bar{T}$ and $Z_{H}t\bar{T}$ couplings equal to zero. For $M=200$
$GeV$ and $500$ $GeV \leq f \leq 1200$ $GeV$, the value of
$\sigma_{1}$ is in the range of $9.4$ $fb \sim 4.6\times10^{-2}$
$fb$, which can generate several and up to hundreds
$t\bar{T}+\bar{t}T$ events per year at the $ILC$ with a yearly
integrated luminosity $\pounds=100$ $fb^{-1}$.

\begin{figure}[htb]
\begin{center}
\epsfig{file=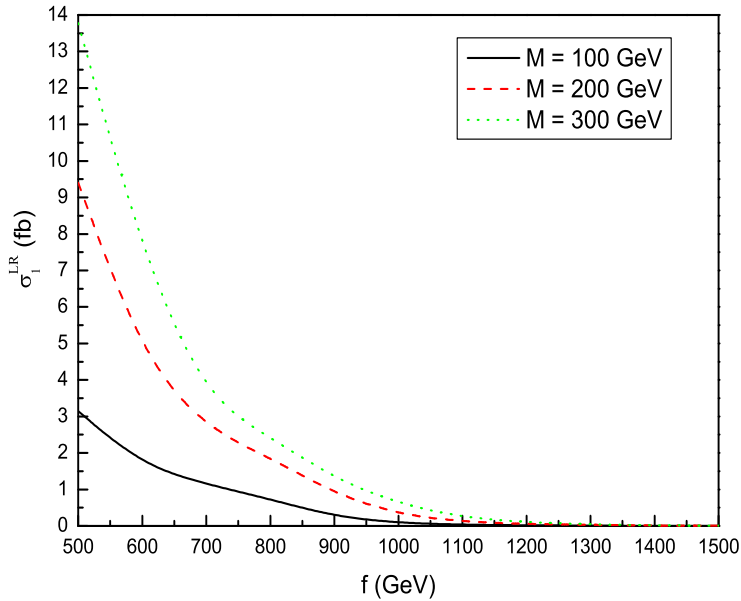,width=320pt,height=280pt}
 \caption{The production cross sections $\sigma_{1}^{LR}$ as a function of the $VEV$
  value $f$ for the \hspace*{1.8cm} $c. m.$ energy $\sqrt{s} = 2$ $TeV$ and three values of the mixing
parameter $M$. We \hspace*{1.9cm}have taken into account both $T$
and $\bar{T}$ quarks.}
\end{center}
\end{figure}

In wide range of the parameter space of the $LRTH$ model, the
possible decay modes of the top partner $T$ are $\phi^{+}b$, $ht$,
$Zt$, $Wb$, and $t\phi^{0}$. It has been shown that, for $M=150$
$GeV$ and $500$ $GeV \leq f \leq 1500$ $GeV$, the branching ratio
$Br$($T\rightarrow\phi^{+}b$) is larger than $70\%$ and the values
for other branching ratios are smaller than $10\%$ [11].
Furthermore, for $M>10$ $MeV$, there is $Br$($\phi^{+}\rightarrow
t\bar{b}$)$\simeq 100\%$. Thus, the dominate decay mode $\phi^{+}b$
 makes the process $e^{+}e^{-} \rightarrow t\bar{T}+ \bar{t}T$ mainly transfers  to
the final state $t\bar{t}b\bar{b}$. For example, for $M=200$ $GeV$ ,
$\sqrt{s} = 2$ $TeV$, and $f=500$ $GeV$ , the production cross
section of the final state $t\bar{t}b\bar{b}$ can reach $7$ $fb$.
There are two kinds of the main backgrounds for this kind of final
state. The first kind is the large $QCD$ backgrounds, which
primarily come from the process $e^{+}e^{-} \rightarrow
t\bar{t}g^{*}$ with the gluon decaying to a $b\bar{b}$ pair. The
second kind is the electroweak backgrounds, of which the dominant
contributions induced by the process $e^{+}e^{-}\rightarrow
Zt\bar{t}$ with the gauge boson $Z$ decaying to a $b\bar{b}$ pair.
Considering the leptonic, semileptonic and fully hadronic decays of
the $t\bar{t}$ system, the main background processes $e^{+}e^{-}
\rightarrow t\bar{t}g^{*}$ and $e^{+}e^{-} \rightarrow t\bar{t}Z$
have been extensively studied in the literature [17, 18]. According
their conclusions, we have to say that it is very difficult to
discriminate the signals generated by the final state
$t\bar{t}b\bar{b}$ from the backgrounds.

 \begin{figure}[htb]
\begin{center}
\epsfig{file=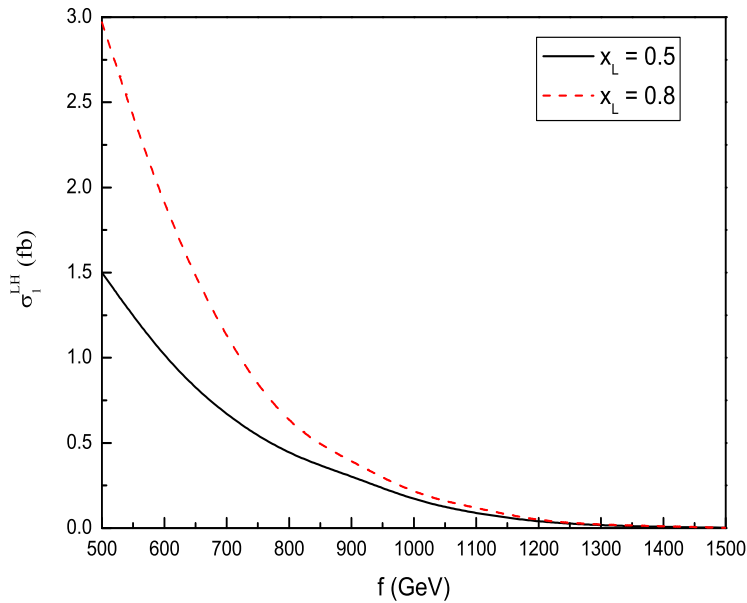,width=320pt,height=280pt}
 \caption{The production cross section $\sigma_{1}^{LH}$ as a function of the scale parameter
  $f$ for \hspace*{1.8cm} the $ c.
m.$ energy $\sqrt{s} = 2$ $TeV$ and two values of the mixing
parameter $x_{L}$. We \hspace*{1.9cm}have taken into account both
$T_{+}$ and $\bar{T}_{+}$ quarks. }
\end{center}
\end{figure}

The second important decay channel of the top partner $T$ is $T
\rightarrow W^{+}b$. However, it is more difficult to separate the
signals generated by the process $e^{+}e^{-} \rightarrow \bar{t}T+ t
\bar{T}$ with $T \rightarrow W^{+}b$ from the large background
coming from the $SM$ process $e^{+}e^{-} \rightarrow t\bar{t}$.
Certainly, the reconstrution of the heavy top partner $T$ through
the combination of $Wb$ may help to distinguish the signals from the
backgrounds. The signals induced by the decay channels $T
\rightarrow th , tZ$, and $t\phi_{0}$, with $h \rightarrow
b\bar{b}$, $Z \rightarrow b\bar{b}$, and $\phi_{0} \rightarrow
b\bar{b}$ can give rise to similar signals with those of the decay
channel $T \rightarrow \phi^{+}b$. However, because of the very
small branching ratios, their production rates are much smaller than
that of the process $e^{+}e^{-} \rightarrow t\bar{T}+ \bar{t}T
 \rightarrow t\bar{t}b\bar{b}$.

The littlest $Higgs$ model with $T$-parity, called the $LHT$ model
[15], also predicts the existence of the top partner $T_{+}$, which
is $T$-even and can also be produced in association with the top
quark $t$ via $e^{+}e^{-}$ collision. The $Feynman$ diagram is
similar with Fig.1. However, the $T$-odd gauge boson $Z_{H}$ can not
contribute the process $e^{+}e^{-} \rightarrow t\bar{T}+ \bar{t}T$.
Our numerical results are given in Fig.3, in which we have plotted
the production cross section $\sigma_{1}^{LH}$ as a function of the
scale parameter $f$ for $\sqrt{s} = 2$ $TeV$ and two values of the
free parameter $x_{L}$. One can see from Fig.3 that, in most of the
parameter space, the value of the cross section $\sigma_{1}^{LH}$ is
smaller than that of the cross section $\sigma_{1}^{LR}$.

The top partner $T_{+}$ can decay into $W^{+}b$, $Ht$, $Zt$, and
$B_{H}t_{-}$, in which $t_{-}$ is the $T$-odd top partner and
$B_{H}$ is the $T$-odd gauge boson. The branching ratios of these
decay modes have been estimated in Ref.[19]. The former three decay
modes can produce the similar signals to those of the top partner
$T$ predicted by the $LRTH$ model. However, their production rates
are smaller than the corresponding production rates in the $LRTH$
model. In most of the parameter space of the $LHT$ model, the
$T$-odd top partner $t_{-}$ mainly decays into $tB_{H}$ and there
are $Br$($t_{-} \rightarrow B_{H}t$) $\approx 100\%$ [19]. If we
assume that $T$-parity is strictly conserved, the lightest $T$-odd
gauge boson $B_{H}$ can be seen as an attractive dark matter
candidate [20]. Than the $B_{H}t_{-}$ decay mode can give rise to
the distinctive state of $t\bar{t}$ plus large missing energy. The
large transverse missing energy can be used to distinguish the
signal from the large $SM$ background $e^{+}e^{-} \rightarrow
t\bar{t}$. However, its production rate is too small to be detected
in the future $ILC$ experiments. Thus, the possible signatures of
top partner predicted by the $LHT$ model or the $LRTH$ model all can
not be detected via the process $e^{+}e^{-} \rightarrow t\bar{T}+
\bar{t}T$ in the future $ILC$ experiments.

\noindent{\bf 4. Single production of the top partner $T$ at future
$ep$ collider}

Although the linac-ring type $ep$ collider ($THERA$) with the
$c$.$m$. energy $\sqrt{s} = 3.7$ $TeV$ and the integral luminosity
$\pounds = 100$ $pb^{-1}$, which is named Energy Frontier $ep$
collider, has a lower luminosity, it can provide better conditions
for studying a lot of phenomena comparing to the $ILC$ due to the
high center-of-mass energy and to the $LHC$ due to more clear
environment [21]. Thus, it is very interesting to consider
production of the new heavy particles at the $THERA$.

From the discussions given in section 2, we can see that the top
partner $T$ can be singly produced via the process $ep \rightarrow
e\bar{b} + X \rightarrow \nu_{e}\bar{T} + X$ at the $THERA$. The
relevant $Feynman$ diagrams are shown in Fig.4. For the subprocess
$e(P_{e})+\bar{b}(P_{b}) \rightarrow
\bar{T}(P_{T})+\nu_{e}(P_{\nu})$, we define kinematical invariants
$\hat{s}=(P_{e}+P_{b})^{2}=(P_{T}+P_{\nu})^{2}$ and
$\hat{t}=(P_{T}-P_{b})^{2}$. The differential cross section is given
by
\begin{eqnarray}
\frac{d\hat{\sigma}_{2}(
\hat{s})}{d\hat{t}}=\frac{\hat{t}^{2}+\hat{t}
(2\hat{s}-M_{T}^{2})+2\hat{s}(\hat{s}-M_{T}^{2})}{64\pi \hat{s}^{2}}
[\frac{(g_{L}^{W\bar{T}b})^{2}(g_{L}^{W\nu_{e}e})^{2}}{(\hat{t}-m_{W}^{2})^{2}}+
\frac{(g_{R}^{W_{H}\bar{T}b})^{2}(g_{R}^{W_{H}\nu_{e}e})^{2}}{(\hat{t}-m_{W_{H}}^{2}
)^{2}}].
\end{eqnarray}

\begin{figure}[htb]
\begin{center}
\epsfig{file=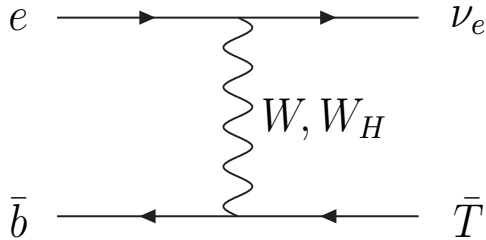,width=180pt,height=90pt}
 \caption{In the $LRTH$ model, the $Feynman$ diagrams for the subprocess $e\bar{b}\rightarrow \nu_{e}\bar{T}$.}
 \label{ee}
\end{center}
\end{figure}

After calculating the cross section $\hat{\sigma}_{2}(\hat{s})$
contributed by the $t$-channel $W$ exchange and $W_{H}$ exchange,
the effective production cross section $\sigma_{2}(s)$ can be
folding $\hat{\sigma}_{2}(\hat{s})$ with the bottom-quark
distribution function $f_{b}(x)$ in the proton
\begin{eqnarray}
\sigma_{2}(s)=\int^{1}_{x_{min}}f_{b}(x,
\mu_{F})\hat{\sigma_{2}}(\hat{s})dx
\end{eqnarray}
with $x_{min}=M_{T}^{2}/s$ and $\hat{s}=xs$. In our numerical
calculation, we will use $CTEQ6L$ parton distribution function [22]
for $f_{b}(x, \mu_{F})$ and assume that the factorization scale
$\mu_{F}$ is of order $\sqrt{\hat{s}}$.

\vspace{0cm}
\begin{figure}[htb]
\begin{center}
\vspace{0.2cm}
 \epsfig{file=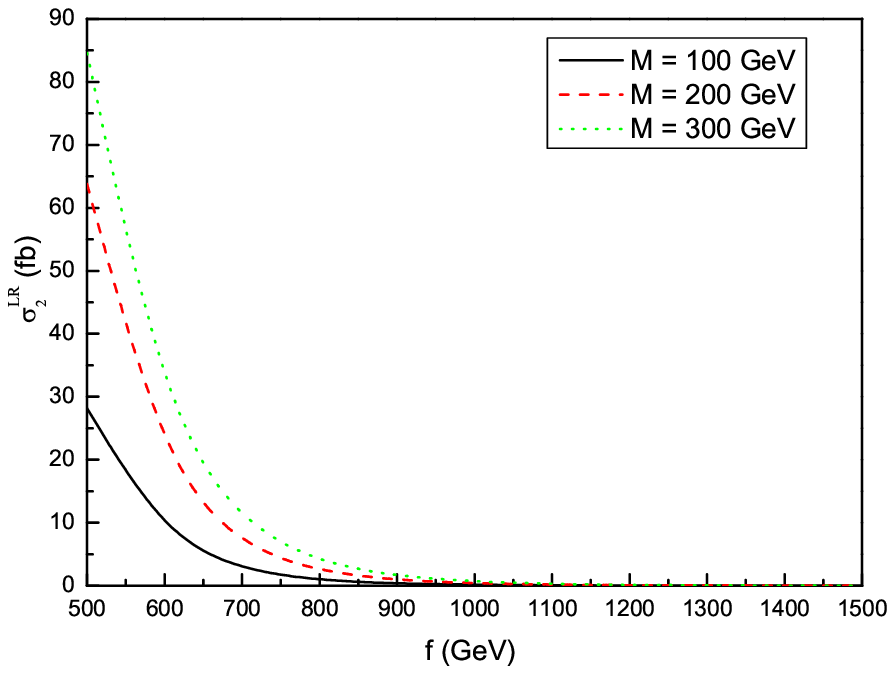,width=245pt,height=220pt}
\put(-130,3){ (a)}\put(80,3){ (b)}
 \hspace{-1.4cm}\vspace{-0.25cm}
\epsfig{file=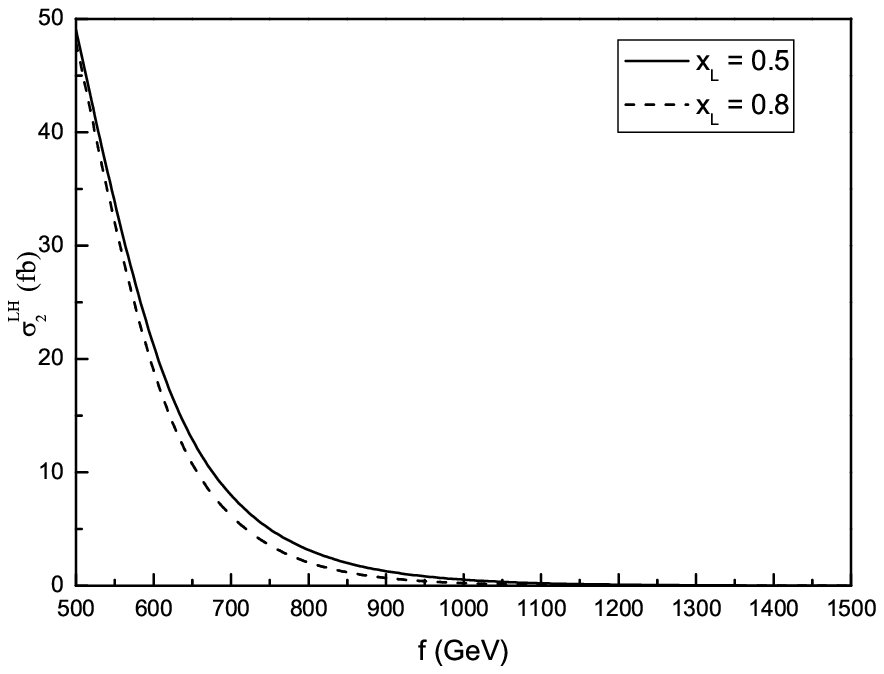,width=245pt,height=220pt} \hspace{-0.5cm}
 \hspace{10cm}\vspace{-1cm}
 \caption{The effective cross sections $\sigma_{2}$ of the subprocess
 $e\bar{b}\rightarrow\ \nu_{e}\bar{T}$ for the $LRTH$ \hspace*{1.8cm} model(Fig.5a) and
 the $LHT$ model(Fig.5b).}
 \label{ee}
\end{center}
\end{figure}

In Fig.5a we plot the cross section $\sigma^{LR}_{2}$ for single
production of the top partner $T$ at the $THERA$ with $\sqrt{s}=3.7$
$TeV$ as a function of the scale parameter $f$  for three values of
the mixing parameter $M$. To compare single production of $T$ at the
$THERA$ with that of the $T$-even top partner $T_{+}$ predicted by
the $LHT$ model, $\sigma_{2}^{LH}$ is shown in Fig.5b as a function
of the scale parameter $f$ for the mixing parameter $x_{L} = 0.5$
and $0.8$. One can see from Fig.5 that the single production cross
section of the top partner at the $THERA$ is larger than that at the
$ILC$. The cross section $\sigma_{2}^{LH}$ is not sensitive to the
free parameter $x_{L}$ and its value is in the range of $47.9$ $fb
\sim 0.19$ $fb$ for $x_{L} = 0.8$ and $500$ $GeV \leq f \leq 1000$
$GeV$. For $100$ $GeV \leq M \leq 300$ $GeV$ and $500$ $GeV \leq f
\leq 1000$ $GeV$, the value of $\sigma_{2}^{LR}$ is in the range of
$72.7$ $fb \sim 0.1$ $fb$. Thus, in most of the parameter space,
$\sigma_{2}^{LR}$ is larger than $\sigma_{2}^{LH}$.

\begin{figure}[htb]
\begin{center}
\epsfig{file=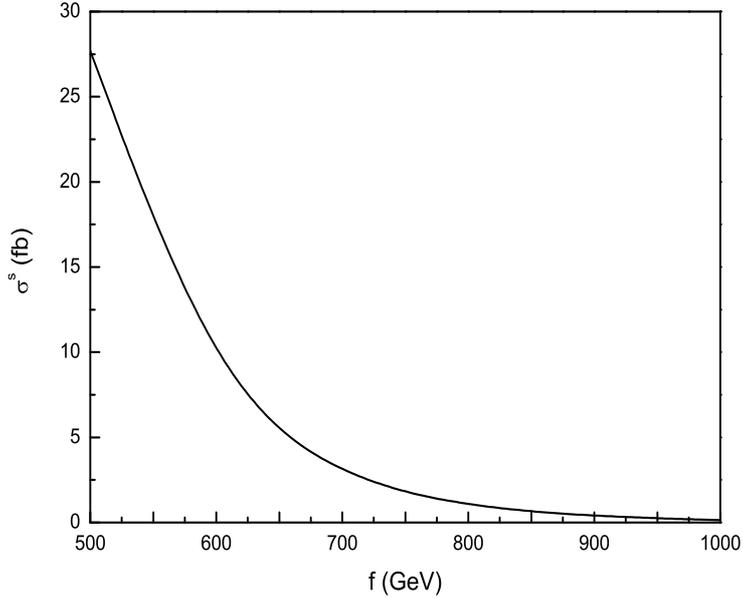,width=320pt,height=280pt}
 \caption{The production rate of the $\nu_{e}\bar{t}b\bar{b}$ final
state as a function of the parameter $f$ for \hspace*{1.8cm}the
mixing parameter $M= 150GeV$.}
\end{center}
\end{figure}

For the dominating decay mode $\phi^{+}b$ of the top partner $T$,
its single production can give rise to the $\nu_{e}\bar{t}b\bar{b}$
final state, which is almost free of the $SM$ background [23]. For
the decay modes $th$, $tZ$, and $t\phi_{0}$, if we assume that there
are $h \rightarrow b\bar{b}$, $Z \rightarrow b\bar{b}$, and
$\phi_{0} \rightarrow b\bar{b}$, the process $ep \rightarrow
\nu_{e}\bar{T}+X$ can also generate $\nu_{e}\bar{t}b\bar{b}$ final
state. The production rate of the $\nu_{e}\bar{t}b\bar{b}$ final
state can be easily estimated $\sigma^{s}\approx
\sigma_{2}\times[Br(T \rightarrow \phi^{+}b) \times
Br(\phi^{+}\rightarrow t\bar{b})+Br(T \rightarrow th) \times
Br(h\rightarrow bb)+Br(T \rightarrow \ tZ) \times Br(Z\rightarrow
b\bar{b})+Br(T \rightarrow \ t \phi^{0}) \times
Br(\phi^{0}\rightarrow b\bar{b})]$. The numerical results are shown
in Fig.6. One can see from this figure that, with reasonable values
of the free parameters of the $LRTH$ model, the production rate can
reach $27fb$. It is obvious that, for the decay channel
$T\rightarrow W^{+}b$, the mainly background of the process
$ep\rightarrow \nu_{e}\bar{T}+X$ comes from the $SM$ process
$ep\rightarrow \nu_{e}\bar{t}+X$, which has been extensively studied
in Ref.[24]. The production rate of this kind of signal  is too
small to be separated from the large background.

For the $T$-even top partner $T_{+}$ predicted by the $LHT$ model,
the decay modes $W^{+}_{H}b$ and $B_{H}t$ make the process
$ep\rightarrow \nu_{e}\bar{T}_{+}$ generate the final state
$\bar{t}+X$. Its dominating background is also the $SM$ process
$ep\rightarrow \nu_{e}\bar{t}+X$. In wide range of the parameter
space of the $LHT$ model, there are $Br(Z_{H}\rightarrow
B_{H}H)\simeq 1$. In the case of $H\rightarrow b\bar{b}$, the decay
modes $Ht$ and $Z_{H}t$ can give rise to similar signal to those of
the decay modes $\phi^{+}b$, $th$, $tZ$ of the top partner $T$.
However, its production rate is much smaller than that generated by
the $LRTH$ model. Thus, the possible signatures of the top partner
$T$ predicted by the $LRTH$ model might be detected via the process
$ep\rightarrow \nu_{e}\bar{T}+X \rightarrow
\nu_{e}\bar{t}b\bar{b}+X$ at the $THERA$, while it is not this case
for the $T$-even top partner $T_{+}$ predicted the $LHT$ model.
\vspace{0cm}

\noindent{\bf 5. Production of the top partner $T$ associated with
the top quark $t$ at the \hspace*{0.7cm}$LHC$}

\begin{figure}[htb]
\begin{center}
\epsfig{file=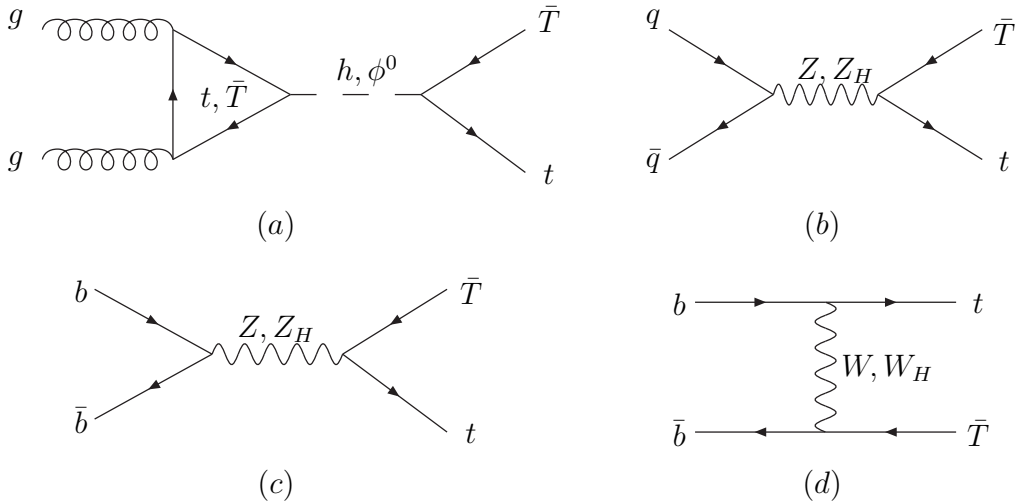,width=380pt,height=190pt}
 \caption{In the $LRTH$ model, the $Feynman$ diagrams for production of the top partner
 \hspace*{1.8cm}$T$ associated with the top quark $t$ at the $LHC$.}
\end{center}
\end{figure}

The $LHC$ will soon go into full operation and provide proton-proton
collisions at a $c.m.$ energy $\sqrt{s}=14$ $TeV$. There are strong
theoretical reasons to expect that the $LHC$ will discover new
physics beyond the $SM$ up to $TeV$. At the $LHC$, the top partner
can be pair-produced via $QCD$ interactions and can be produced
associated with a jet mediated by the $SM$ gauge bosons, which have
been extensively studied in the literature [7, 9, 11, 25]. The top
partner $T$  can also be produced associated with the top quark $t$
via the gluon fusion with neutral scalar exchanges and quark
antiquark annihilation with the gauge boson exchanges. In the $LRTH$
model, the $Feynman$ diagrams for this kind of production processes
are shown in Fig.7.

\vspace{0cm}
\begin{figure}[htb]
\begin{center}
\vspace{0.2cm}
 \epsfig{file=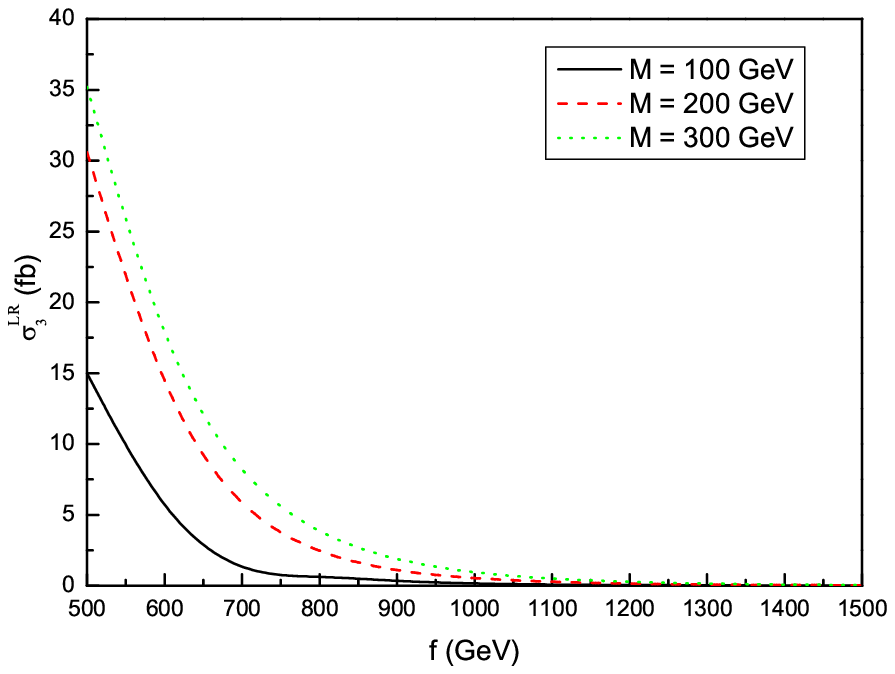,width=245pt,height=220pt}
\put(-130,3){ (a)}\put(80,3){ (b)}
 \hspace{-1.4cm}\vspace{-0.25cm}
\epsfig{file=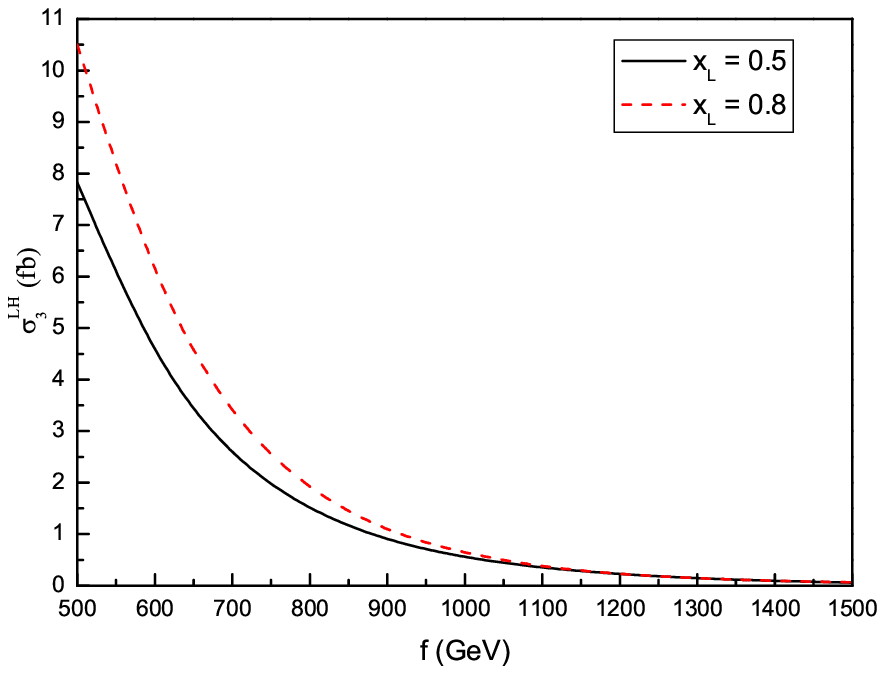,width=245pt,height=220pt} \hspace{-0.5cm}
 \hspace{10cm}\vspace{-1cm}
 \caption{The hadronic cross sections $\sigma_{3}$ for the subprocess
 $gg\rightarrow\ t\bar{T}+\bar{t}T$ as function of \hspace*{1.8cm}the parameter $f$ in the
 $LRTH$ model (a) and the $LHT$ model (b).}
 \label{ee}
\end{center}
\end{figure}

\vspace{0cm}
\begin{figure}[htb]
\begin{center}
\vspace{0.2cm}
 \epsfig{file=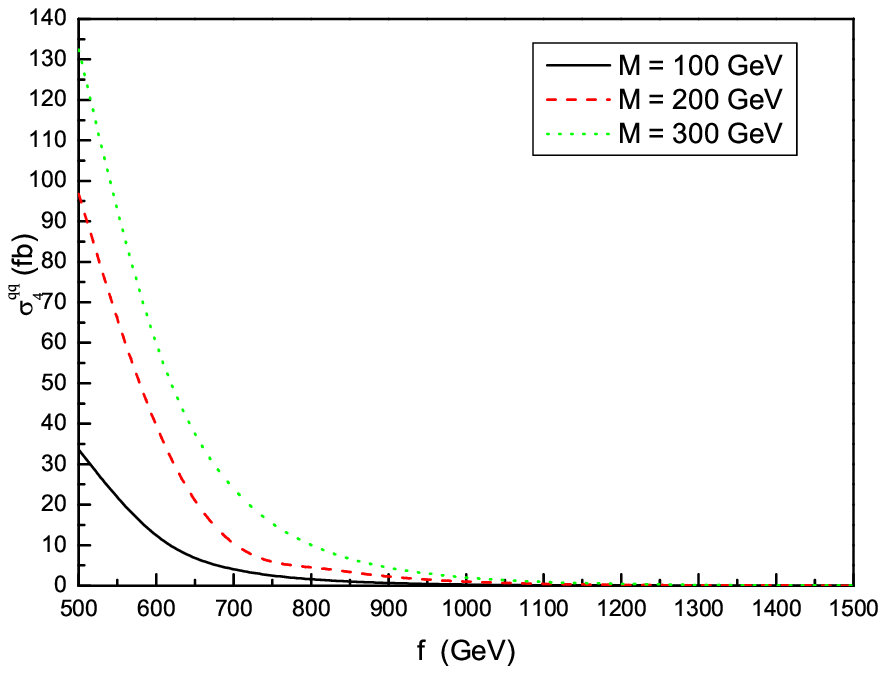,width=245pt,height=220pt}
\put(-130,3){ (a)}\put(80,3){ (b)}
 \hspace{-1.4cm}\vspace{-0.25cm}
\epsfig{file=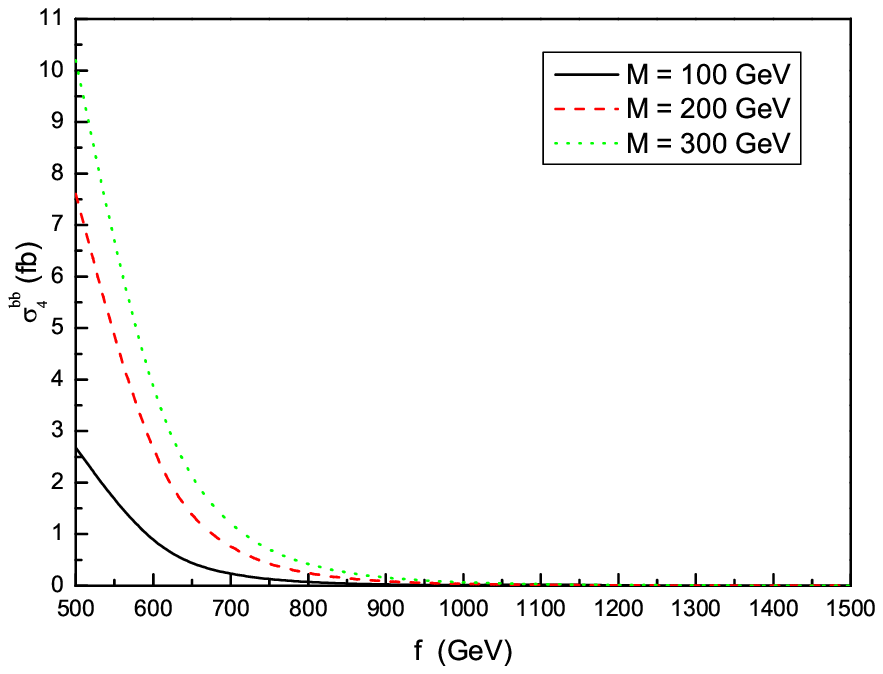,width=245pt,height=220pt} \hspace{-0.5cm}
 \hspace{10cm}\vspace{-1cm}
 \caption{The hadronic cross sections $\sigma_{4}$ for the
 subprocesses
 $q\bar{q}\rightarrow\ t\bar{T}+\bar{t}T$(a) and \hspace*{1.8cm}$b\bar{b}\rightarrow\
 t\bar{T}+\bar{t}T$(b) as a function of the parameter $f$ for three
 values of the mixing  \hspace*{1.8cm}parameter $M$.}
 \label{ee}
\end{center}
\end{figure}

Similar with sections 3 and 4, using the relevant $Feynman$ rules we
can easily give the expressions of the parton level cross sections
for the subprocesses $q\bar{q}\rightarrow t\bar{T}+\bar{t}T$ ($q=u$,
$c$, $d$, and $s$) and $b\bar{b}\rightarrow t\bar{T}+\bar{t}T$. For
the gluon-induced production of the top partner as shown in Fig.7a,
we first obtain the effective couplings $hgg$ and $\phi^{0}gg$
induced by the fermion loop, which have been extensively studied in
the literature [26], then the parton level cross section
$gg\rightarrow t\bar{T}+\bar{t}T$ can be easily calculated. It is
well known that the hadronic cross section can be obtained by
folding the parton level cross section with the $PDFs$ in the
proton. Our numerical results are summarized in Fig.8 and Fig.9, in
which Fig.8 and Fig.9 correspond to the subprocesses $gg\rightarrow
t\bar{T}+\bar{t}T$ and $q\bar{q}\rightarrow t\bar{T}+\bar{t}T$
($q=u$, $c$, $d$, $s$, and $b$), respectively. The $T$-even top
partner $T_{+}$ predicted by the $LHT$ model can also be produced in
association with the top quark $t$ at the $LHC$. The relevant
$Feynman$ diagrams are similar with those of Fig.7. However, the new
scalars and new gauge bosons predicted by the $LHT$ model are
$T$-odd, which have no contributions to the subprocesses
$gg\rightarrow t\bar{T}+\bar{t}T$ and $q\bar{q}\rightarrow
t\bar{T}+\bar{t}T$. Thus, its production cross section induced by
the gluon fusion mechanism is smaller than that of the top partner
$T$ as shown in Fig.8b. In most of the parameter space of the $LHT$
model, the value of the hadronic cross section induced by the
subprocess $q\bar{q}\rightarrow t\bar{T}+\bar{t}T$ is smaller than
$5\times10^{-2}$ $fb$, so the relevant curve lines are not shown in
Fig.9. One can see from these figures that, in the $LRTH$ model, the
hadronic cross section from the subprocess $q\bar{q}\rightarrow
t\bar{T}+\bar{t}T$ is larger than that from the subprocess
$gg\rightarrow t\bar{T}+\bar{t}T$, while the contributions of the
subprocess $b\bar{b}\rightarrow t\bar{T}+\bar{t}T$ to the hadronic
cross section should not be neglected. For $100$ $GeV\leq M \leq300$
$GeV$ and $500$ $GeV\leq f \leq1500$ $GeV$, the value of the total
hadronic cross section is in the range of $177.6$ $fb\sim0.013$
$fb$. If we assume the yearly integrated luminosity
$\pounds_{int}=100$ $fb^{-1}$ for the $LHC$ with the $c.m.$ energy
$\sqrt{s}=14$ $TeV$, then there will be several up to ten thousands
of the $t\bar{T}+\bar{t}T$ events to be generated per year.

\begin{figure}[htb]
\begin{center}
\epsfig{file=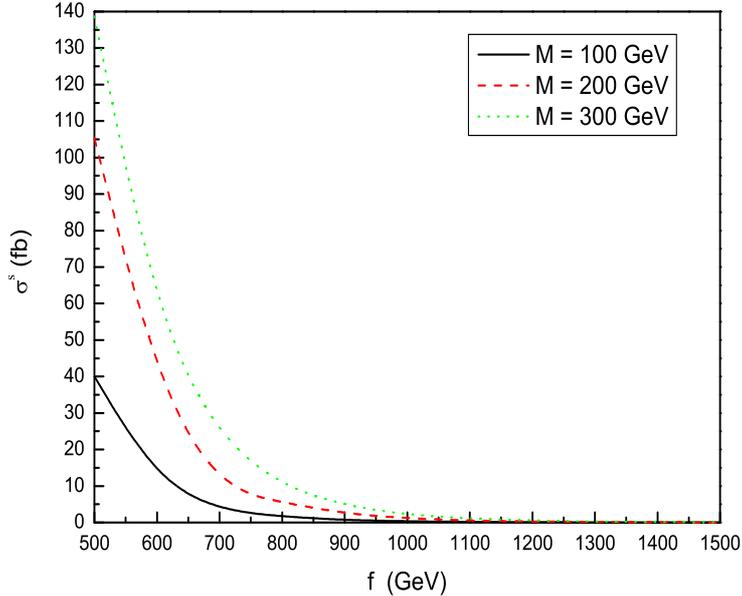,width=320pt,height=280pt}
 \caption{In the $LRTH$ model, production rate for the
$t\bar{t}b\bar{b}$ final state as a function of  \hspace*{1.8cm}the
parameter $f$ for three values of the mixing parameter $M$.}
\end{center}
\end{figure}

From the discussions given in section 3, the decay modes
$\phi^{+}b$, $ht$, $Zt$, and $\phi^{0}t$ of the top partner $T$ with
$h\rightarrow b\bar{b}$, $Z\rightarrow b\bar{b}$ and
$\phi^{0}\rightarrow b\bar{b}$ can make the process $pp\rightarrow
t\bar{T}+\bar{t}T+X$ give rise to the $t\bar{t}b\bar{b}$ final
state. Its production rate is plotted in Fig.10 as a function of the
parameter $f$ for three values of the mixing parameter $M$. One can
see that, with reasonable values of the free parameters, the
production rate can be larger than $100fb$. However, its value
decreases quickly as the parameter $f$ increasing. The main
backgrounds for the $t\bar{t}b\bar{b}$ final state come from the
$SM$ processes $PP\rightarrow$ $t\overline{t}Z$ + $X$ and
$PP\rightarrow t\overline{t}h + X$ with $Z\rightarrow b\overline{b}$
and $h\rightarrow b\overline{b}$, continuum $t\overline{t}
b\overline{b}$ production, and the reducible background
$t\overline{t}+jets$, in which the additional jets light quarks or
gluons but may be misidentified as $b$ quarks. Detailed analysis of
the signals and the relevant backgrounds have been given in
Ref.[27]. They have shown that the most large background
$t\overline{t}+jets$ can be suppressed by enhancing the detector
ability to tag $b$ quark jets. Furthermore, It may be help to
distinguish the $t\bar{t}b\bar{b}$ final state from the second large
background coming from the process $PP\rightarrow t\overline{t}h +
X$ by calculating the invariant mass of the $tb\bar{b}$ state
generated by the top partner $T$ decaying. Thus, it may be possible
to extract the signals from the backgrounds by applying suitable
cuts. Certainly, detailed confirmation of the observability of the
 signals generated by the process $pp\rightarrow
t\bar{T}+\bar{t}T+X$, would require Monte-Carlo simulations of the
signals and backgrounds, which is beyond the scope of this paper. It
is obvious that it is more difficult to separate the signals
generated by the process $pp\rightarrow t\bar{T}+\bar{t}T+X$ with $T
\rightarrow W^{+}b$ from the large background coming from the $SM$
process $pp\rightarrow t\bar{t}+X$. Certainly, the reconstrution of
the heavy top partner $T$ through the combination of $Wb$ may help
to distinguish the signals from the backgrounds.

\noindent{\bf 6. Conclusions}

The twin $Higgs$ mechanism provides an alternative method to solve
the little hierarchy problem. The $LRTH$ model is a concrete
realization of the twin $Higgs$ mechanism, which predicts the
existence of the top partner $T$. In this paper, we consider
production of $T$ associated with the top quark $t$ in future $ILC$
and $LHC$ experiments, and its single production in future $THERA$
experiments. To compare our results obtained in the context of the
$LRTH$ model with those of the $LHT$ model, we also estimate
production of the $T$-even top prater $T_{+}$ via the corresponding
processes in these high energy collider experiments. From our
numerical results, we can obtain the following conclusions.

i) In most of the parameter space, the production  cross sections of
the top partner $T$ are larger than those of the $T$-even top
partner $T_{+}$ in these three kinds of collider experiments.
However, all of their values decrease quickly  as the parameter $f$
increasing.

ii) In the context of the $LHT$ model, the production  cross section
of the process $pp\rightarrow t\bar{T}+\bar{t}T+X$ is larger than
that of the process $e^{+}e^{-}\rightarrow\ t\bar{T}+\bar{t}T$ or
the process $ep\rightarrow\ \nu_{e} \bar{T}+X$. For $100$ $GeV\leq M
\leq300$ $GeV$ and $500$ $GeV\leq f \leq1500$ $GeV$, the value of
the production cross section for the process $pp\rightarrow
t\bar{T}+\bar{t}T+X$ is in the range of $177.6$ $fb\sim0.013$ $fb$.

iii) The decay modes of the top partner $th , tZ$, and $t\phi_{0}$,
with $h \rightarrow b\bar{b}$, $Z \rightarrow b\bar{b}$, and
$\phi_{0} \rightarrow b\bar{b}$ can give rise to similar signals
with those of the decay modes $ \phi^{+}b$. They can produce the
$\nu_{e}\bar{t}b\bar{b}$ and $t\bar{t}b\bar{b}$ final states in the
$THERA$ and $LHC$ experiments, respectively. Although the production
rate of the $\nu_{e}\bar{t}b\bar{b}$ final state is smaller than
that of the $t\bar{t}b\bar{b}$ final state, the
$\nu_{e}\bar{t}b\bar{b}$ final state is more easy detected because
of it almost free of the $SM$ backgrounds [23]. Considering the very
large backgrounds, the decay channel $T \rightarrow W^{+}b$ can not
be used to detecting the possible signals of the top partner $T$ in
future high energy collider experiments.

\vspace{1.5cm}
 \noindent{\bf Acknowledgments}

This work was supported in part by the National Natural Science
Foundation of China under Grants No.10675057, Specialized Research
Fund for the Doctoral Program of Higher Education(SRFDP)
(No.200801650002), the Natural Science Foundation of the Liaoning
Scientific Committee(No.20082148), and Foundation of Liaoning
Educational Committee(No.2007T086). 
\end{document}